\documentclass[12pt]{article}
\usepackage{times}

\newcounter{lastnote}

\usepackage{color}

\title{Observation of squeezed light with 10\,dB quantum noise reduction}

\author{Henning Vahlbruch, Moritz Mehmet, Nico Lastzka, Boris Hage, Simon Chelkowski,\\ Alexander Franzen, Stefan Go\ss ler, Karsten Danzmann and Roman Schnabel $^{\ast}$\\
\\
\normalsize{Max-Planck-Institut f\"ur Gravitationsphysik
(Albert-Einstein-Institut) and}\\ \normalsize{Institut f\"ur
Gravitationsphysik, Leibniz Universit\"at Hannover}\\
\normalsize{Callinstr. 38, 30167 Hannover, Germany}\\
\\
\normalsize{$^\ast$To whom correspondence should be addressed;
E-mail:  Roman.Schnabel@aei.mpg.de.} }

\date{\today}

\usepackage{graphicx}

\begin{document}


\baselineskip24pt

\maketitle


\textbf{Squeezing of light's quantum noise requires temporal rearranging of photons. This again corresponds to
creation of quantum correlations between individual photons. Squeezed light is a non-classical manifestation
of light with great potential in high-precision quantum measurements, for example in the detection of
gravitational waves \cite{Cav81}. Equally promising applications have been proposed in quantum communication
\cite{YSh78}. However, after 20 years of intensive research doubts arose whether strong squeezing can ever be
realized as required for eminent applications. Here we show experimentally that strong squeezing of light's
quantum noise is possible. We reached a benchmark squeezing factor of 10 in power (10\,dB). Thorough analysis
reveals that even higher squeezing factors will be feasible in our setup. 
}

Theoretical considerations about the possible existence of light with squeezed quantum noise can be traced
back to the 1920's. However, only after applications for squeezed light were proposed in the 1980's squeezing
was discussed in more detail  \cite{YSh78,Hol79,Cav81,Wal83, Dod02}. 
In \cite{Cav81} it
was suggested to use squeezed light to improve the sensitivity of kilometre-scale Michelson laser-interferometers
for the detection of gravitational waves.
Proof of principle experiments have been successfully conducted \cite{MSMBL02,VCHFDS05} and squeezed states have been generated
also in the audio signal band of ground-based detectors \cite{MGBWGML04,VCHFDS06}.
Another field of application is \emph{continuous variable} (CV) quantum communication and information
\cite{YSh78,BLo05}. While \emph{discrete variable} quantum information typically relies on single photon
detectors, which are limited in terms of detection speed and quantum efficiency, squeezed light is detected
with homodyne and heterodyne detectors which reveal quantum correlations by averaging over a vast number of
detected photons. Due to this, high bandwidth and almost perfect detection efficiencies are possible.
Squeezed states of light have been used to demonstrate several CV quantum information
protocols. They have been used to construct entangled states of light and to demonstrate quantum
teleportation \cite{FSBFKP98,BTBSRBSL03,TYAF05}. They are a possible resource for secure quantum key
distribution protocols \cite{GCe06,NGA06} and for generation of cluster states for universal quantum
computing \cite{MLGWRN06}. Recently, squeezed states of light have been used to prepare Schr\"odinger kitten
states for quantum information networks \cite{OTLG06,NNHMP06}.

For all proof of principle experiments so far only modest strengths of squeezing were available. In
\cite{MSMBL02,VCHFDS05,MGBWGML04,VCHFDS06} about 3 to 4\,dB of squeezing was achieved.
The first CV teleportation experiments \cite{FSBFKP98, BTBSRBSL03} did not reach the so-called no-cloning
limit of fidelity greater than 2/3 \cite{GGr01} due to the limitations in squeezing strength.
Although the first experimental demonstration of squeezed light succeeded in 1985 \cite{SHYMV85}, dedicated
research in the following two decades could only elaborate typical factors of 2 to 4 (3\,dB to 6\,dB), see
also \cite{BSM97,TGBFBL03}. However, very recently a great step forward was achieved at the University of
Tokyo and a factor of 8 (9\,dB) quantum noise squeezing of a laser field at 860\,nm was observed
\cite{TYYF07}. This wavelength is close to atomic transitions having important implications for quantum
information storage \cite{HHGLBBL06}.
In our experiment we generated a squeezed laser beam with a quantum noise reduction of a factor of 10 at a laser wavelength of 1064\,nm which is used in current gravitational wave detectors \cite{AD05}.


\begin{figure}[t!]
\centerline{
\includegraphics[width=12cm]{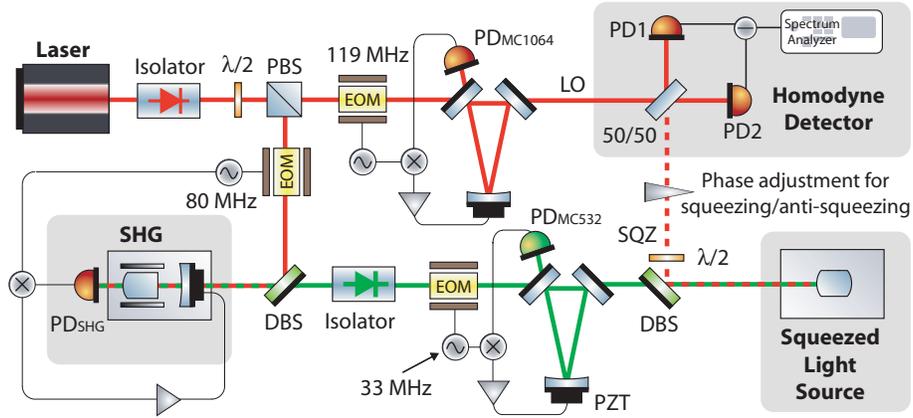}}
  \vspace{0mm}
\caption{{\bf{A}} Schematic of the experimental setup. Squeezed states of light (SQZ) at 1064\,nm were generated by
type\,I optical parametric oscillation (OPO) below threshold. SHG: second harmonic generation, PBS:
polarizing beam splitter;  DBS: dichroic beam splitter; LO: local oscillator, PD: photodiode; EOM:
electro-optical modulator.}
  \label{experiment}
\end{figure}

As shown in Fig.\,\ref{experiment} the laser source of our experiment was a monolithic non-planar Nd:YAG ring
laser of 2\,W continuous wave single mode output power.
Approximately 1.9\,W were used for second harmonic generation (SHG) to provide the pump field at 532\,nm for
our optical parametric squeezed light source. A detailed description of the SHG design can be found in
\cite{CVDS07}.
An important feature of our experiment were two travelling-wave resonators which served as optical low-pass
filters for phase noise on the laser beams as well as spatial mode cleaners. These cavities were positioned
in the beam path of both the fundamental and second harmonic field; one cavity close to the homodyne detector
and one close to the squeezed light source. Both resonators had a finesse of 350 and  a linewidth of
1.44\,MHz. The cavities were held on resonance with the laser fields via a Pound-Drever-Hall locking scheme.
These resonators significantly reduced phase front
mismatches and phase fluctuations. It has been shown in \cite{TYYF07, FHDFS06} that phase fluctuations, for
example of the second-harmonic pump field, can be a limiting factor for 
strong squeezing.

Our squeezed light source was a monolithic cavity made from 7\,\% doped MgO:LiNbO$_{3}$ that produced
squeezed states via type\,I degenerate optical parametric oscillation (OPO. The crystal
length was 6.5\,mm and both front and rear face had a radius of curvature of 8\,mm. Each surface was
dielectrically coated to give power reflectivities of 88\,\% or 99,97\,\% at 1064\,nm, respectively.
Second harmonic pump powers between 650\,mW and 950\,mW were mode-matched into the squeezed light source and parametric gains between 63 to more than 200 were observed. Squeezed states
were produced when the crystal temperature was stabilized at its phase-matching temperature 
and the laser wavelength was tuned on resonance with the squeezed light source cavity. Due to the high stability
of our setup no  servo-loop control for the laser frequency was required. The squeezed states
left the source in counter direction of the pump field and were separated 
via a dichroic beam splitter (DBS). The observation of (squeezed) quantum noise was performed by means of a balanced homodyne detector built from a pair of Epitaxx ETX-500 photodiodes. We achieved a fringe visibility of 99.8\,\% between the squeezed beam and the local oscillator on
the 50/50 homodyne beam splitter. 

\begin{figure}[t!]
  \centerline{\includegraphics[width=10cm]{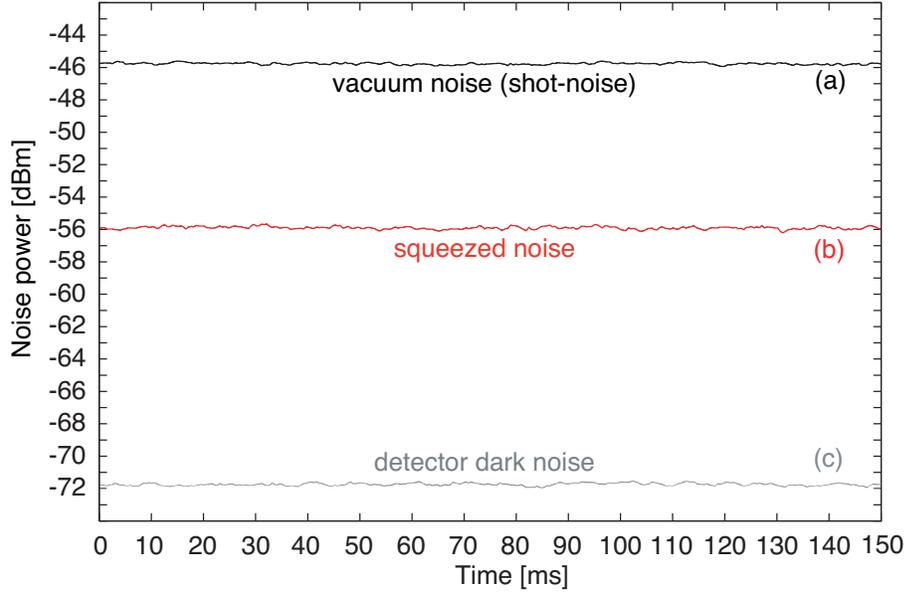}}
  \vspace{0mm}
\caption{Quantum noise powers at a Fourier frequency of 5\,MHz, measured with a resolution bandwidth of
100\,kHz and video bandwidth of 100\,Hz. Trace (a) shows the vacuum noise level corresponding to 26.9\,mW
local oscillator power. Trace (b) shows the noise power of the squeezed vacuum states measured with the same
local oscillator power. A nonclassical noise reduction of 10.12\,dB below vacuum noise was observed. The
electronic detector dark noise is shown in trace (c) and was not subtracted from the data. Each trace was
averaged three times.}
  \label{squeezing}
\end{figure}

Fig.\,\ref{squeezing} presents the first ever direct observation of light with 10\,dB squeezing. Shown are noise powers
at the Fourier sideband frequency of 5\,MHz. Trace (a) corresponds to the shot-noise of uncorrelated photons
of 26.9\,mW local oscillator power and was measured with the squeezed
light input blocked. In this arrangement no photons entered the signal port of the homodyne detector and the
measured shot-noise can be directly linked to the vacuum noise, which corresponds to the light's quantum
mechanical ground state. Trace (b) shows the quantum noise-reduction when squeezed states were injected. The directly observed squeezing level was 10.12 ($\pm$ 0.15)\,dB. The detector dark noise (trace (c)) was approximately 26\,dB below the vacuum noise level. Darknoise subtraction leads to a squeezing level of 10.22 ($\pm$ 0.16)\,dB.

\begin{figure}[t!]
\centerline{\includegraphics[width=10cm]{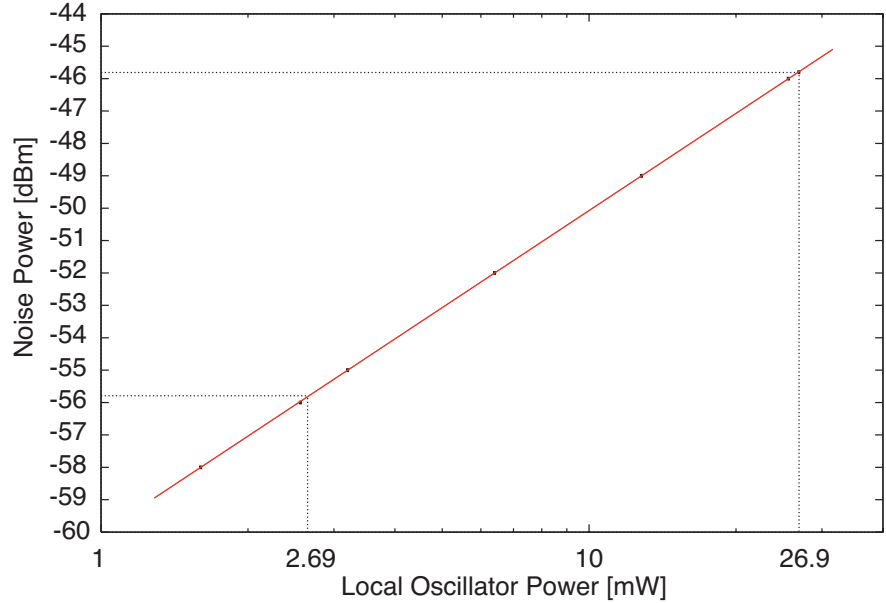}}
  \vspace{0mm}
\caption{The linearity of the homodyne detection system  was validated by varying the local oscillator power. 
Shown is the linear fit to seven measurement values (squares).
The sizes of the squares corresponds to the measurement error bars. 
Note that the shot-noise of a laser beam of 2.69\,mW is shown to be identical to the squeezed noise of the differential mode in our homodyne detector with ten times the light power, compare with figure 1.
}
  \label{LOPower}
\end{figure}

\begin{figure}[t!]
\centerline{\includegraphics[width=10cm]{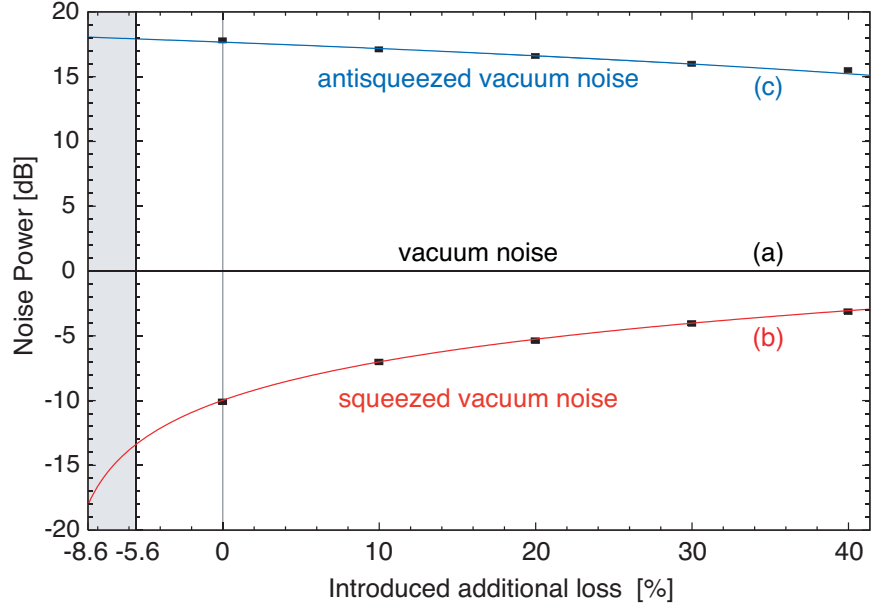}}
  \vspace{0mm}
\caption{Squeezing and anti-squeezing levels for a parametric gain of 63 versus optical loss. Solid lines
show the theoretical predictions. Square boxes represent measurement values with sizes corresponding to the errors bars. Electronic darknoise was subtracted in this figure. The two vertical axis on the left corresponds to the upper and lower boundaries of how much squeezing might be achieved in our setup by reduction of optical loss.}
  \label{LossChart}
\end{figure}

To confirm the observed squeezing strength, we checked linearity of the homodyne detection system
including the spectrum analyser by measuring shot-noise levels versus local oscillator powers
(Fig.\,\ref{LOPower}). A linear fit matches the measurements accurately.
To further validate the observation of 10\,dB squeezing we introduced a known amount of optical loss into the
squeezed light beam. The observed squeezing and anti-squeezing strength should depend on this additional loss in a
characteristic way. For this procedure a combination of a $\lambda/2$ waveplate and a polarizing
beam splitter was placed between the 50/50 beam splitter of the homodyne detector and each photodiode (PD1,\,PD2). 
Since both fields -- the squeezed beam and the local oscillator -- suffered from the loss, the intensity of the local oscillator
beam was re-calibrated to the nominal value of 26.9\,mW by using a more intense beam in front of the modecleaner.
Fig.\,\ref{LossChart} shows the observed amount of squeezing and anti-squeezing with an additional 10\%, 20\%, 30\%, and 40\% introduced optical loss, respectively. The solid lines (b) and (c) represent the simulations for a parametric gain of $g=63$ which was experimentally realized with 650\,mW pump power. We found excellent agreement with the experimental data.


With an increased pump power of 950\,mW we observed anti-squeezing of 23.3\,dB whereas the squeezing was
still 10\,dB below vacuum noise. 
This observation can be used to deduce boundaries for the total optical loss in our setup. 
Assuming a loss free setup in which the observed squeezing strength is limited by anti-squeezing coupling into our squeezing measurement via phase fluctuations, we derived the upper limit for phase jitter to be $\phi=1.2^\circ$.
Since $\phi$ is independent of the pump power we can conclude that 10\,dB squeezing, as observed with 650\,mW (and less anti-squeezing), was not limited by phase fluctuations but optical loss.  Even with $\phi=1.2^\circ$ we find the minimum value for the total optical loss in our setup to 5.6\,\%. Secondly we assumed phase fluctuations of $\phi \ll 1.2^\circ$. Here the observed squeezing is completely limited by optical loss, which results in the upper bound of 8.6\,\%. 
Taking these boundaries into account, the left part of Fig.\,\ref{LossChart} shows how much squeezing might be achieved in our setup by optical loss reduction. 

In independent measurements we
determined the intra-cavity round trip loss of the squeezed light source at 1064\,nm to be less than 0.07\%,
corresponding to an escape efficiency of the squeezed states from the source in excess of 99.4\%. Loss during
propagation occurs due to the dichroic beam splitter and non-perfect anti-reflection coatings of lenses and
were determined to be about 1.1\%. The non-perfect visibility at the homodyne beam splitter introduced
another 0.4\% of loss. Given these values we estimate the quantum efficiency of the ETX-500 photodiodes to be 95($\pm$2)\,\%. 
Our analysis suggests that the non-perfect quantum efficiency of our photodiodes was the main limitation in
our experiment. With improved photodiodes close to unity quantum efficiency, which already exist for shorter
wavelengths \cite{TYYF07}, an additional factor of 2 in quantum noise reduction might be possible.

The direct observation of 10\,dB squeezing of quantum noise of light, as reported here, shows that the
squeezed light technique has indeed a great application potential as envisaged more than two decades ago. Injected into a
gravitational wave detector, the quantum noise reduction corresponding to an increase of factor 10 in laser
light power will be possible \cite{Cav81}. This is a promising application, since gravitational wave
detectors already use the highest single-mode laser powers applicable. Furthermore, our results might enable the generation of strongly entangled states to reach teleportation
fidelities well above 2/3 as already typically achieved in single photon teleportation experiments
\cite{teleportation}.

\section*{Acknowledgement}
We would like to thank H.-A. Bachor, W. P. Bowen, J. Harms, S. Hild, N. Mavalvala, D. E. McClelland, K.
McKenzie, P. K. Lam, and A. Th\"uring  for helpful discussions regarding the optimization of squeezed light
generation and detection. This work has been supported by the Deutsche Forschungsgemeinschaft and is part of
Sonderforschungsbereich 407.



\end{document}